\documentclass[proof]{pasj01}
\draft

\newcommand\bs{\boldsymbol}
\newcommand\rad{\mathrm{rad}}

\begin{document}
\SetRunningHead{Author(s) in page-head}{Running Head}
\Received{}
\Accepted{}

\title{Observations of Photospheric Magnetic Structure 
below a Dark Filament 
using the Hinode Spectro-Polarimeter }

\author{
Takaaki \textsc{Yokoyama}\altaffilmark{1},
Yukio \textsc{Katsukawa}\altaffilmark{2,3},
Masumi \textsc{Shimojo}\altaffilmark{2,3}
}
\altaffiltext{1}{Department of Earth and Planetary Science, The University of Tokyo, \\
  7-3-1 Hongo, Bunkyo-ku, Tokyo 113-0033}
\altaffiltext{2}{National Astronomical Observatory of Japan, NINS,
  2-21-1 Osawa, Mitaka, Tokyo 181-8588}
\altaffiltext{3}{Department of Astronomical Science, SOKENDAI
(The Graduate University of Advanced Studies), Mitaka, Tokyo, 181-8588, Japan}


\maketitle

\begin{abstract}
The structure of the photospheric vector magnetic field below a dark filament 
on the Sun is studied using the observations of the Spectro-Polarimeter
attached to the Solar Optical Telescope onboard {\it Hinode}. 
Special attention is paid to discriminate the two suggested models, 
a flux rope or a bent arcade.  ``Inverse-polarity'' 
orientation is possible below the filament in a flux rope, whereas
``normal-polarity'' can appear in both models. We study a filament in
active region NOAA 10930, which appeared on the solar disk during 
2006 December. 
The transverse field perpendicular to the line of sight
has a direction almost parallel 
to the filament spine with a shear angle of 30 deg, whose orientation
includes the 180-degree ambiguity. To know whether it is 
in the normal orientation or in the inverse one, the center-to-limb 
variation is used for the solution under the assumption that
the filament does not drastically change its magnetic structure during the
passage.
When the filament is near the east limb, 
we found that the line-of-site magnetic component below is positive, 
while it is negative near the west limb. 
This change of sign indicates 
that the horizontal photospheric field perpendicular to the 
polarity inversion line beneath the filament 
has an ``inverse-polarity'', which indicates a flux-rope structure
of the filament supporting field.
\end{abstract}

\section{Introduction}

Dark filaments or prominences are one of the most impressive objects observed
by the optical H$\alpha$ spectral line in the solar atmosphere
(reviews in, e.g., 
\cite{parenti2014, mackay2010, tandberghanssen1995, hirayama1985}).
Cold heavy matter stays above the surface at an altitude of up to 100 Mm
surrounded by hot coronal plasma. 
As suggested by the occurrence above the magnetic polarity inversion lines
(PILs; \cite{babcock1955}),
they are considered to be supported
by the magnetic field opposing the gravity.

Numbers of theoretical studies 
have been conducted on the magnetic structure of 
filaments. 
\citet{kippenhahn1957}
suggested that the cool material is
stored at the dips near the top 
of the bent arcade fields because 
of its own weight. In another model proposed by 
\citet{kuperus1974},
the filament
material is stored at the lower dips 
inside a flux rope above the PIL. 
In the Kippenhahn-Schl\"uter model, the transverse magnetic component
crossing the filament 
has a ``normal'' polarity, i.e. the orientation of the field is
from the positive to the negative local vertical polarities while
it is in the opposite direction, i.e., an ``inverse'' polarity, 
in the Kuperus-Raadu model. 

Studying the equilibrium structure is not only important by itself but
also is interesting for understanding of a filament eruption, which
is frequently followed by a large flare and a coronal mass ejection.
In some observations, a slow evolution is found before an eruption 
\citep{feynman1995, chifor2006, nagashima2007, isobe2006}.
This is considered to be a manifestation of the approaching process 
to the de-stabilization or the loss of equilibrium.
In the model by 
\citet{forbes1991}
the equilibrium state
evolves slowly in response to, e.g., the surface motion
(e.g., \cite{kaneko2014}).
Through this evolution, the magnetic pressure force of the fields
below a flux rope finally overcomes the tension force of the overlying magnetic
field lines that is pulling the rope back. 
\citet{chen2000}
extended this idea that
the final de-stabilization is triggered by an emerging flux in the
vicinity of a flux rope corresponding to a filament. 
In both these models, the final state just before an eruption
has a flux rope in the corona. 
On the other hand, models in which the eruption appears 
from an arcade structure are also proposed.
The stretching motion along the PIL with a
converging motion enhances the
magnetic pressure by the stretched axial field 
(\cite{amari2000})
leading to the trigger of an eruption.  
The observation of a filament's
magnetic structure is necessary  to understand
the feasibility of these proposed processes.

The magnetic fields of the prominences have been
diagnosed using the spectro-polarimetric data of the optical spectral lines
(A review of the early works 
is found in 
\cite{leroy1989}).
They are the results of 
the inversion based on the Hanle or Zeemann effects.
The early observations revealed that the magnetic field has a
strength of around 10 G and is mostly horizontal and makes an acute angle
of about 40 deg relative to the spine 
(\cite{leroy1989, bommier1998};
reviews in 
\cite{mackay2010}).
\citet{casini2003}
obtained the first map of the vector field of a prominence.
They found organized patches of 
magnetic field significantly stronger (up to 80 G)
in the surrounding weak horizontal field.

To infer the field structure of a filament,
the measurement of the photospheric magnetic field is an indirect
method but relatively more precise than those in the filament itself.
Using the detailed polarimetric measurements, 
\citet{lites2005}
reconstructed the photospheric vector
magnetic field below some narrow active region filaments in comparison
with high resolution H$\alpha$ observations.
He found a flux rope structure
in the filaments, i.e., an inverse polarity structure of the photospheric field.
\citet{lopezariste2006}
also found bald patch structures in the bipoles in a 
filament channel as well as at the foot of its barbs. 
\authorcite{okamoto2008}(\yearcite{okamoto2008,okamoto2009})
found an evolving magnetic field
underneath an active region filament and interpreted that its
evolution is consistent with an emerging flux rope structure.
\citet{kuckein2012}
conducted a multiwavelength multiheight study
of the vector magnetic field of an active region filament and found that
its inferred fields suggest a flux rope topology.

This study reports the magnetic structure beneath
a filament in the photosphere that is
observed by the Spectro-Polarimeter  attached with the Solar
Optical Telescope  onboard the {\it Hinode} spacecraft, that is
capable of observing the four Stokes components of Fe I 6301.5 \& 
6302.5 \AA\ doublet
lines emanating from the photosphere. These spectro-polarimetric 
data are fitted by our developed code, MEKSY,
based on the Milne-Eddington atmospheric model. 
As for the practical procedure of the measurement of the transverse
magnetic component, there remains a fundamental difficulty
of the 180-degree uncertainty.
The approach described in this study uses the 
center-to-limb evolution of the Stokes-V signal to solve the 180-degree
ambiguity in the azimuthal angle of the photospheric vector
field below the filament (see also 
\cite{bommier1981}).
This method assumes that
the global magnetic structure of the filament is static during this period.

Based on the measurement in this study, 
we attempt to discuss the global structure of
the filament.
In the Kippenhahn-Schl\"uter bent-arcade model, 
the transverse magnetic component
crossing or beneath the cool material has normal polarity.
In the Kuperus-Raadu flux-rope model, the crossing field has inverse polarity.
However, for the photospheric field,
there are two possibilities for the orientation:
First, if the flux rope is detached above the photosphere, 
low-height loops can connect in a normal manner, and the
magnetic field close to the photosphere can have normal orientation.
Second, if part of the flux rope field lines cross the photoshpere,
a so-called bald patch configuration appears close to the PIL
in the photosphere, and an inverse orientation can be observed.

In the next section, the observation is described. The results are
given in section 3 with corresponding discussions.

\section{Observation and Analysis}

\begin{figure}
\begin{center}
\includegraphics[width=160mm]{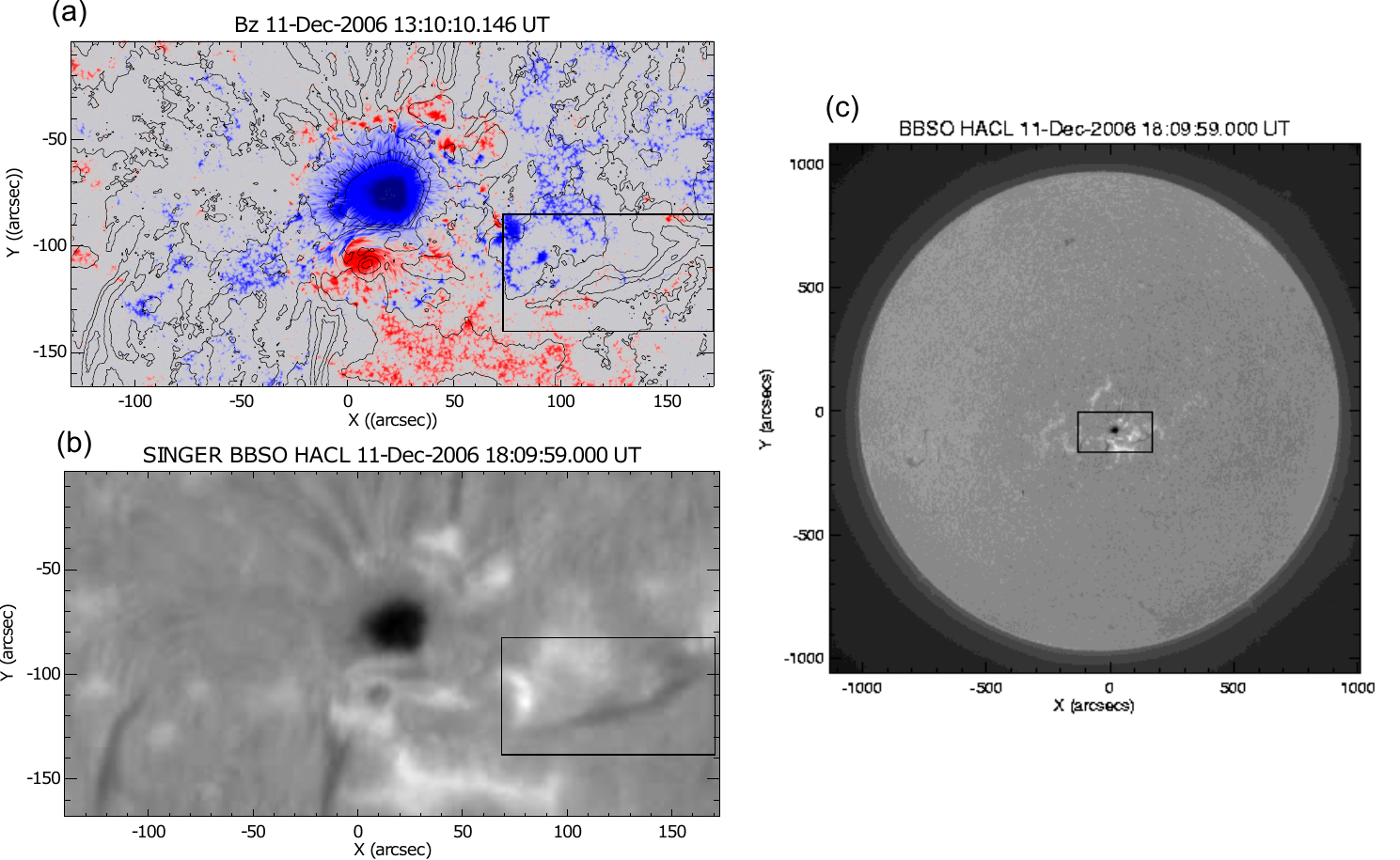}
\end{center}
\caption{Active region NOAA 10930 on 2006 December 11. Color map in (a): 
LOS component of the magnetic field obtained by the Milne-Eddington
fitting procedure from the {\it Hinode} SOT/SP 
observations. Red (blue) color indicates the 
positive (negative) polarity.
The color plot is saturated at an absolute strength at 3 kG
with dark blue (red) for negative (positive) field.
Contours in (a) and gray-scale in (b) : Brightness in the H$\alpha$
band taken at the Big Bear Solar Observatory. The square insets 
in panels (a) and (b) 
indicates the filament studied in this paper.
(c) The same data as (b) but in full observational field of view. The square line 
indicates the field of view of panels (a) and (b)
}\label{d061211_xysp_filament_f}
\end{figure}

We used the data sets taken by the observations of active region NOAA 10930
(figure \ref{d061211_xysp_filament_f})
during its passage on the solar disk from 2006 December 8 to 16.
The square insets of panels a and b in figure \ref{d061211_xysp_filament_f}
indicate the analyzed filament.

We used full-sun images in the hydrogen H$\alpha$ band taken by the
Global High-Resolution H$\alpha$ Network which consists of facilities
at different global longitudes 
\citep{steinegger2001}
as information on the chromospheric geometry 
of the filaments.
In the analyzed period, data from the Big Bear Solar Observatory 
(BBSO; figure \ref{d061211_xysp_filament_f}b and c)
and Meudon Observatory are available. We also used data
taken by the Solar Magnetic Activity Research Telescope (SMART) 
at the Hida observatory of Kyoto University
\citep{ueno2004}
and by the Polarimeter for Inner Coronal Studies (PICS)
of the Advanced Coronal Observing System (ACOS) at Mauna Loa Solar Observatory.
The spatial
resolution depends on the capability of the telescope for each image
and also on the seeing level at the data acquisition. 
Unfortunately,
the achieved resolutions are
insufficient to resolve the fine structures in the chromospheric fibrils. 

We analyzed the Stokes profiles obtained by the Spectro-Polarimeter (SP;
\cite{lites2013})
attached 
with the Solar Optical
Telescope (SOT; 
\cite{tsuneta2008, suematsu2008, ichimoto2008, shimizu2008})
on board the {\it Hinode} spacecraft 
\citep{kosugi2007}
for the photospheric magnetic field.
The SP obtains the profiles of two magnetically sensitive Fe I lines at
6301.5 \AA\ and 6302.5 \AA. Stokes IQUV data are obtained
with a polarimetric accuracy of $< 0.1\%$.
The sampling pixels are 0.15 arcsec $\times$ 0.16 arcsec in the normal-map mode 
and 0.30 arcsec $\times$ 0.32 arcsec in the fast-map mode.
The spectral resolution (sampling) is 30 m\AA\ (21.5 m\AA).
The field of view along the slit is 163.84 arcsec in the 
solar north-south direction. The single exposure duration is 4.8 sec 
at one slit position.  The cadence
between the maps is determined by the trade-off with telemetry and
memory storage. We have one to five maps per day during the 
analyzed period. 
The SOT/SP maps are coaligned with H$\alpha$ images
by using their continuum images by visual inspection.
Both SP continuum and H$\alpha$ images contain a large main sunspot
of the active region, which helps this alignment procedure. The
precision of the alignment is mostly limited by the spatial resolution,
i.e. the seeing quality, of the H$\alpha$ images.

The Stokes profiles measured by the SP 
are processed with a standard calibration routine 
SP\_PREP 
\citep{litesichi2013}
available under the SolarSoft package, and they
are fitted by a spectral model
to obtain the physical parameters in the
solar photosphere. The fitting model is based on the Milne-Eddington 
atmosphere in which the physical parameters, such as magnetic field
strength, orientation, and Doppler velocity, are assumed to be homogeneous
along the line of sight. An exact analytical solution is available 
as a set of explicit formulae known as the Unno-Rachkovsky solution
(\cite{unno1956, rachkovsky1962a}; \yearcite{rachkovsky1962b}).
The non-linear fitting based on this solution 
is conducted using the MEKSY code. The details regarding this
code are provided in Appendix \ref{secappendix2}.
Note that MEKSY itself does not take care of the so-called ``180-degree
ambiguity'' in the azimuth angle of the magnetic field.
The resolution of this ambiguity is the heart of this analysis 
and will be described in the next section and Appendix
\ref{secappendix1} in detail.

In this study, we use the heliocentric Cartesian coordinate with
$x$, $y$, and $z$. The
origin is at the solar center, and $x$ and $y$ are, respectively,
in the westward and northward directions 
and $z$ is in the earthward direction.
Further, we use heliographic coordinates $r$, $\Theta$, and $\Phi$, where
$\Theta$ is the latitude and $\Phi$ is the longitude.
$\Phi=0$ corresponds to the terrestrial observer's central meridian.

\section{Results and Discussion}

\begin{figure}
\begin{center}
\FigureFile(125mm,92mm){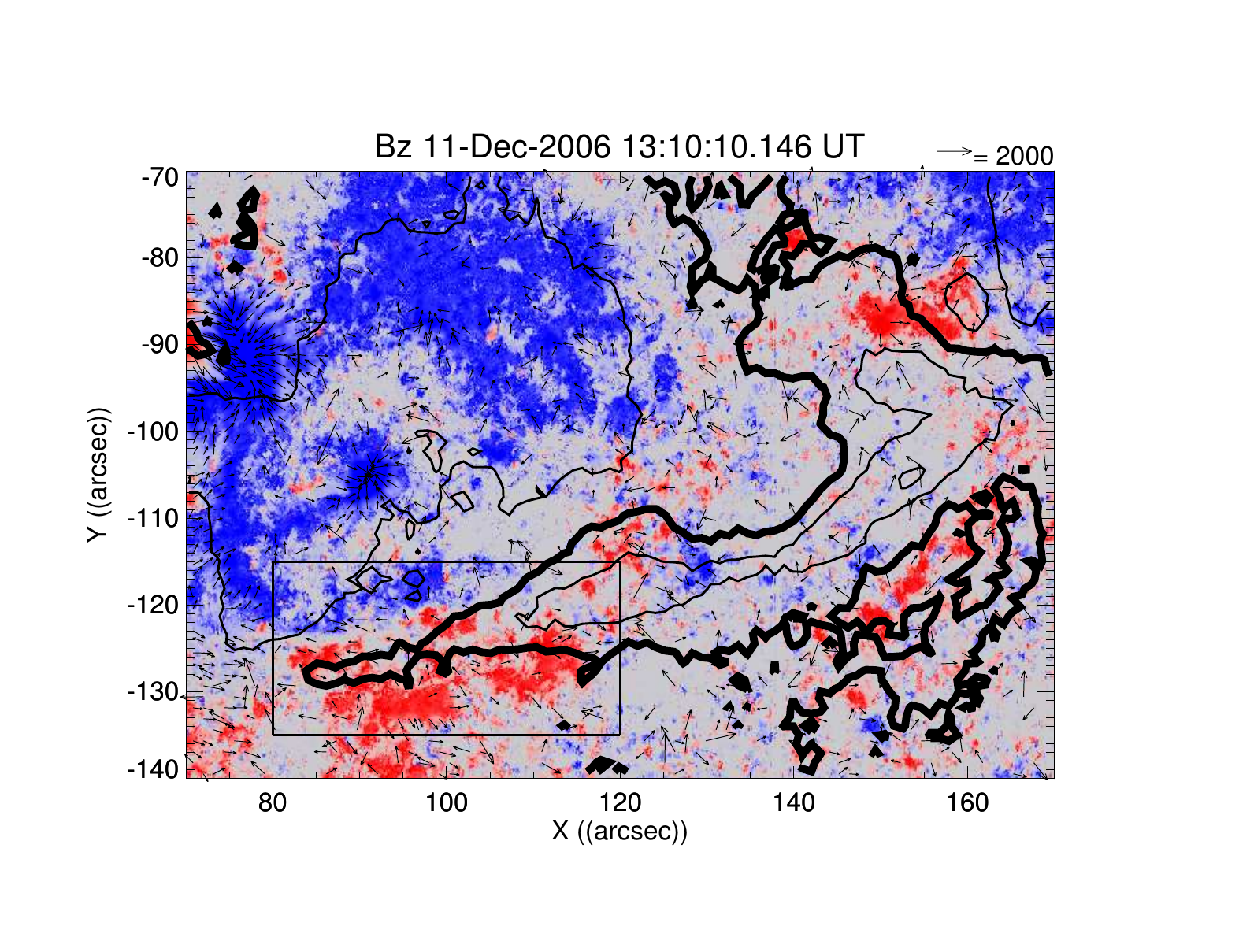}
\end{center}
\caption{LOS (color) and transverse (arrows) components of 
the magnetic field obtained by the Milne-Eddington
fitting procedure from the {\it Hinode} SOT/SP observations.
The color plot is saturated at an absolute strength at 3 kG
with dark blue (red) for negative (positive) field.
Note that the orientation of the transverse components is determined
by the component closest to the extracted potential field based on the observed
LOS component distribution. 
Solid contours are for brightness in the H$\alpha$ 
band taken at the Big Bear Solar Observatory. 
The thick contour line indicates the outline of the analyzed filament.
The solid box indicates the region where the analysis was performed.
}\label{d061211_xysp_filament2_f}
\end{figure}
%
\begin{figure}
\begin{center}
\FigureFile(125mm,92mm){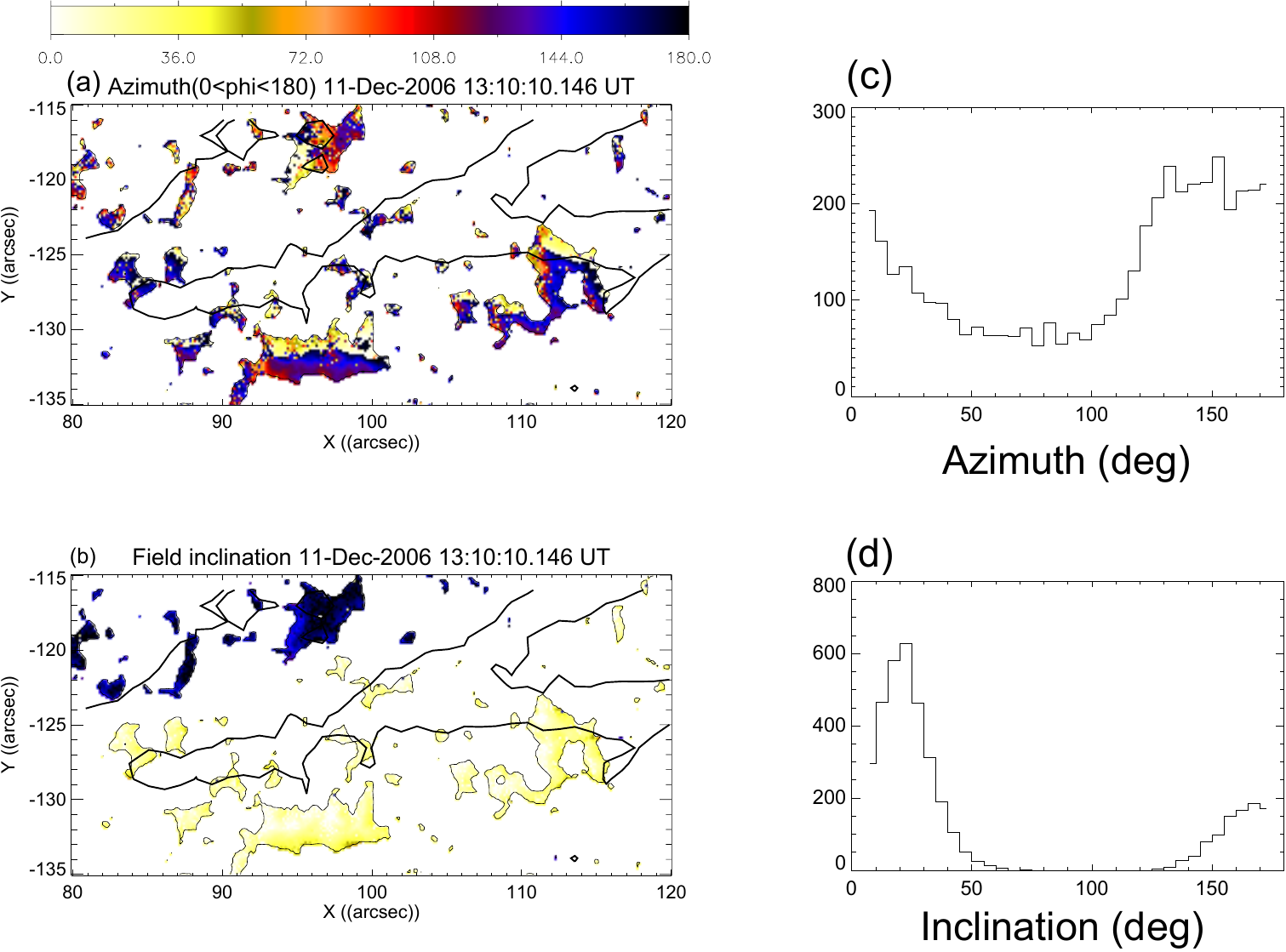}
\end{center}
\caption{(a) Azimuth angle $\chi_B$ and (b) inclination $\gamma_B$
of the magnetic field obtained by the Milne-Eddington fitting procedure
from the {\it Hinode} SOT/SP observations on 2006 Dec. 11. 
The azimuth is defined by the
angle between the transverse component of the magnetic field
and the heliocentric $x$-axis in the anti-clockwise direction.
Because of the 180-degree ambiguity, the orientation of the
transverse component is indistinguishable.
The inclination is defined by the angle between the
earthward direction and the magnetic field.
The analyzed area is shown 
as a solid box in figure \ref{d061211_xysp_filament2_f}.
Histograms of (c) the azimuth angle of the transverse components
and (d) the inclination angle of the magnetic field.
Note the strong peak in the bin for a value of zero include the
data points of the weak polarization degree ($< 1\%$).
}\label{d061211_xysp_azinc}
\end{figure}

The square insets of figure \ref{d061211_xysp_filament_f} indicate the analyzed 
filament. The filament is located at the outer edge of the active region
and its lifetime is longer than the duration of the disk passage
of the active region. The heliographic coordinate on 2006 December 11 is
S07W07 (i.e., $\Theta=-7\deg$ and $\Phi= 7\deg$).
Although several intense flares including those above 
the GOES X-class occurred in this active region, a part of this filament
maintained its appearance during the analyzed period 
as an east-west oriented dark structure at a position 
2 deg south and 5 deg west of the main sunspot
i.e., $\Theta \approx -9\ \deg$ 
and $\Phi \approx 12\ \deg$
in the H$\alpha$ images 
obtained with a cadence of, at least, one image per day.
Because of this relatively long lifetime at least more than
two weeks,
this filament is considered to have a quiescent nature
despite its close location to the edge of the active region. 
It is shown at the time in figure \ref{d061211_xysp_filament_f}
that the filament is located 
at the boundary between the positive and negative
line-of-sight (LOS) magnetic polarities.
In contrast to this global structure, the detailed magnetic distribution
below the scale less than $10^{4}\ {\rm km}\ (\approx 15\ {\rm arcsec})$ 
has a patchy structure where most of the magnetic flux density is around
200 Mx cm$^{-2}$ (figure \ref{d061211_xysp_filament2_f}).
We limited the analysis on these relatively strong patches concentrated
in the eastern segment of the filament indicated
by the solid box in figure \ref{d061211_xysp_filament2_f}.
The spine of the studied filament is tilted by 
$\chi_f \approx 10\ \deg $ 
toward northwest from the east-west direction.

\begin{figure}
\begin{center}
\includegraphics[width=80mm]{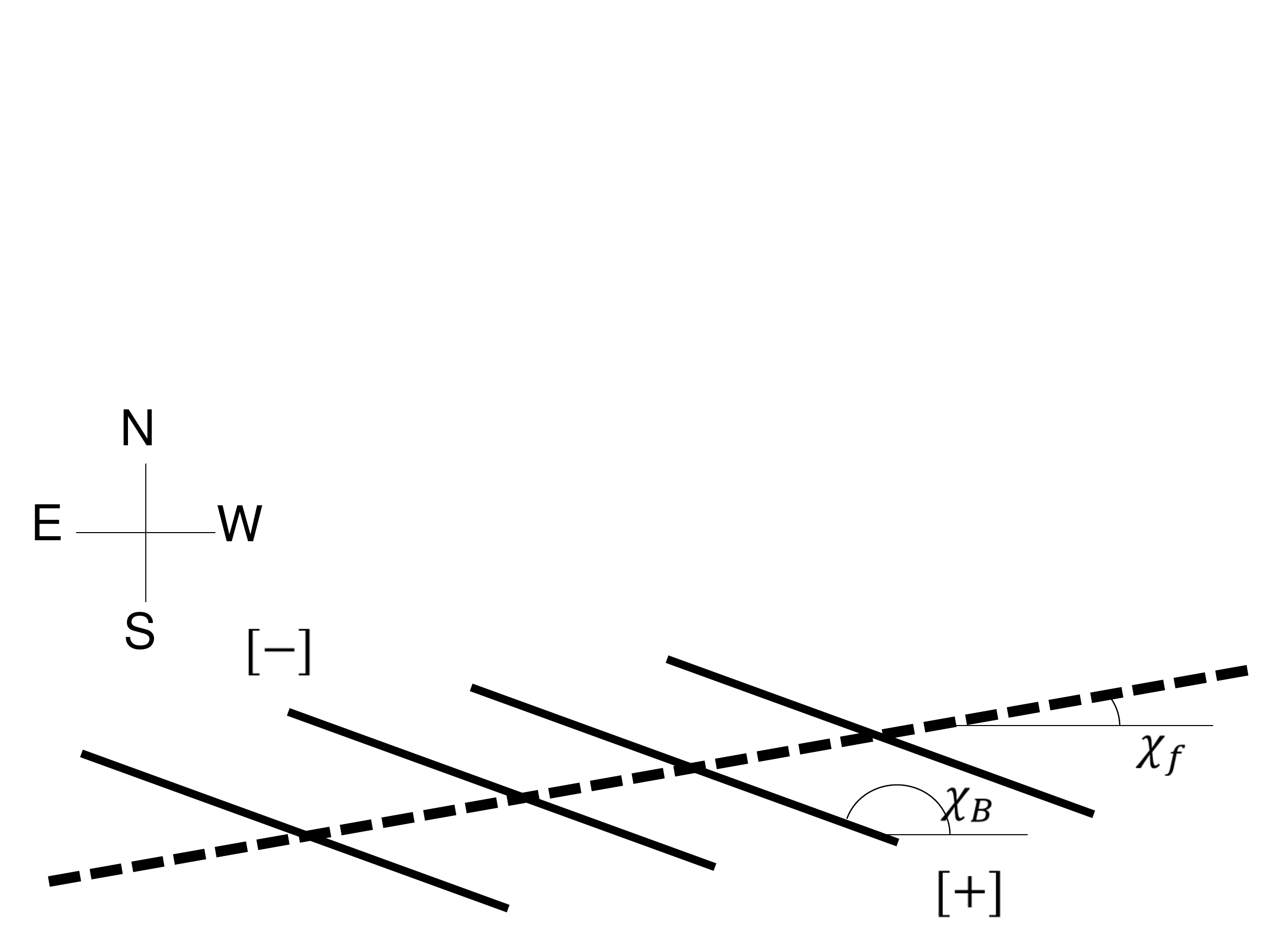}
\includegraphics[width=40mm]{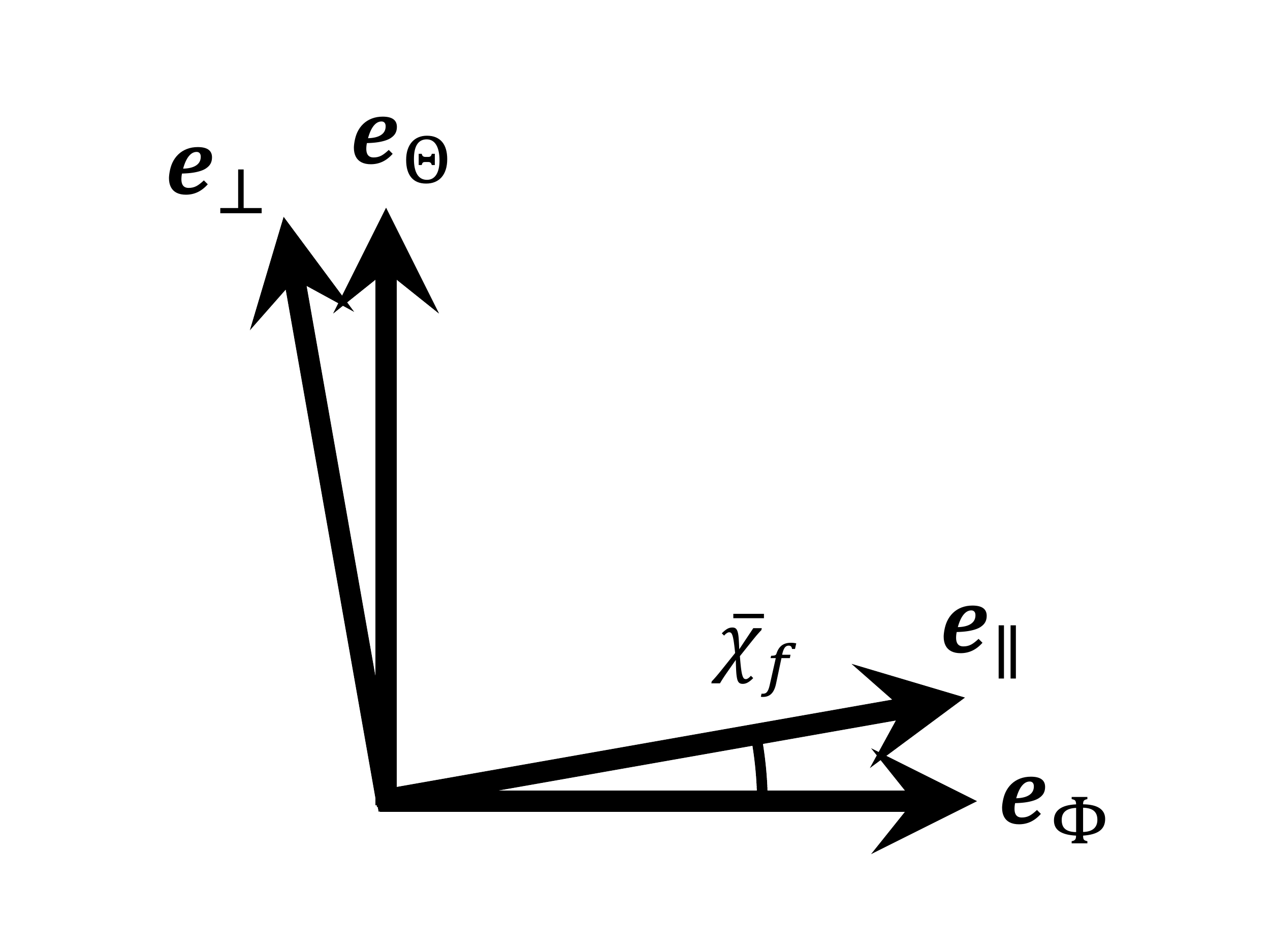}
\end{center}
\caption{Drawing showing the relative angle and location between the
filament and the magnetic field. Dashed line indicates the location
of the PIL which is along the filament
with $\chi_f \approx 10\ \deg$ tilt from the east-west direction.
The direction along the PIL in the
solar surface is indicated by an element vector $\bs{e}_\parallel$ and
the direction perpendicular to the PIL in the same plane
by $\bs{e}_\perp$.
The solid lines indicate 
the photospheric
magnetic field with the azimuth angle $\chi_B$ 
obtained by SOT/SP. Note that the 180-deg
ambiguity remains in the azimuth angle. 
The letters ``$[+]$'' and ``$[-]$'' indicate
the sign of the line-of-site component of the magnetic field.
}\label{figure_p1}
\end{figure}

Figures \ref{d061211_xysp_azinc}a and c show the spatial distribution
and histogram of the azimuth angle $\chi_B$ 
of the transverse component to the LOS of the magnetic field on 2006 Dec. 11. 
$\chi_B$ is the anti-clockwise angle
from the western direction ($x$ direction). Note that only the data points 
with a strong polarization
degree ($> 1\%$) are shown and that the values are
$0 \le {\chi}_B<180\ {\rm deg}$
because  the orientation of the transverse component is indistinguishable
owing to the 180-degree ambiguity.
The distribution peak is around 
${\chi}_B \approx 160\ \deg$. 
The relative angle between this direction and the spine 
of the filament is 
$\chi_B - \chi_f \approx 150\ \deg$
or 30 deg in an acute angle,
i.e. the field is strongly sheared 
(figure \ref{figure_p1}).
This is consistent with the previous statistical results on 
the mid-latitude quiescent prominences by 
\citet{leroy1984}.

To obtain information on the three-dimensional 
magnetic structure of the filament,
the orientation of the transverse component perpendicular to the
PIL beneath the filament is a key parameter.
Assume that the spine of the filament is parallel to the PIL
on the surface of the sun and take an elemental vector 
$\bs{e}_\parallel$ to be parallel to the PIL (figure \ref{figure_p1}).  Then,
\begin{equation}
\begin{array}{cl}
\bs{e}_\parallel&=\bs{e}_\Phi \cos{\bar{\chi}_f}+\bs{e}_\Theta \sin{\bar{\chi}_f},
\\
\bs{e}_\perp    &=-\bs{e}_\Phi \sin{\bar{\chi}_f}+\bs{e}_\Theta \cos{\bar{\chi}_f},
\end{array}
\end{equation}
where $\bar{\chi}_f$ is the tilt angle of the PIL against the latitudinal line.
As shown in figures \ref{d061211_xysp_azinc}b and d,
the north-eastern and south-western sides of the filament have the negative 
($\gamma_B\approx 160\ \deg$, where $\gamma_B$ is an inclination angle 
of the magnetic field) 
and positive ($\gamma_B\approx 25\ \deg$) vertical polarities, respectively. 
Therefore, if the orientation of
the transverse component perpendicular to the PIL is 
$B_\perp<0$ ($B_\perp>0$, see figure \ref{figure_p1}), 
the transverse field has a ``normal  polarity'' (``inverse polarity''). 
For this solution, 
it is necessary to solve the 180-degree azimuth ambiguity. 
We used the apparent change  of the magnetic signals during the rotational 
passage on the solar disk of this filament. 
Because this filament
shows steady appearance during the passage on the solar disk, we assume that
the magnetic structure does not change much during this period.
We hereafter assume that the components $B_r$, $B_\Theta$, and $B_\Phi$
in the heliographic coordinate of the photospheric magnetic field
do not change during the disk passage.
In addition, it is important to note that, owing to the stable 
observational condition on the orbit out of the atmospheric influence, 
{\it Hinode} can take maps with a uniform and stable quality 
for this analyzed region.

In terms of the photospheric magnetic field on the PIL,
the radial component is zero,
$B_r=0$.
For this field,  the line-of-site component $B_z$ and
the transverse component perpendicular to the PIL $B_\perp$ has 
a relation 
\begin{equation}
\frac{B_z}{B_\perp}
\approx
-\frac{ \sin{\Phi}}{\tan\chi_{B0}},
\label{eq050}
\end{equation}
where $\chi_{B0}$ is an observed azimuth angle of the magnetic field
at the PIL when it is located at the disk center.
We used an approximation that $|\Theta|\ll 1 \ \rad$, 
$|\chi_f|\ll 1 \ \rad$, and $|\tan{\chi_B}|\ll 1$ for the derivation of this formula
(see appendix \ref{secappendix1} for a detailed derivation).
Note that $\tan{{\chi}_{B0}}$ is independent of the 180-degree
ambiguity. A crude explanation to this formula is that, the analyzing
PIL (and filament as well) is almost in the east-west direction
($|\chi_f|\ll 1 \ \rad$). Thus $B_\perp$ is almost in the north-south direction
with a slight tilt, namely  $|\tan\chi_B|\ll 1$.  Due to this tilt, when they
are located close to the limb, one can observe the LOS component $B_z$
corresponding to this $B_\perp$. 
In the analyzing filament, $\chi_B\approx 160\ \deg$, then
$\tan\chi_B<0$. Therefore, from eq. (\ref{eq050}), $B_z$ has the same sign with $B_\perp$
in the western hemisphere $\Phi >0$.

\begin{figure}
\begin{center}
\includegraphics[width=130mm]{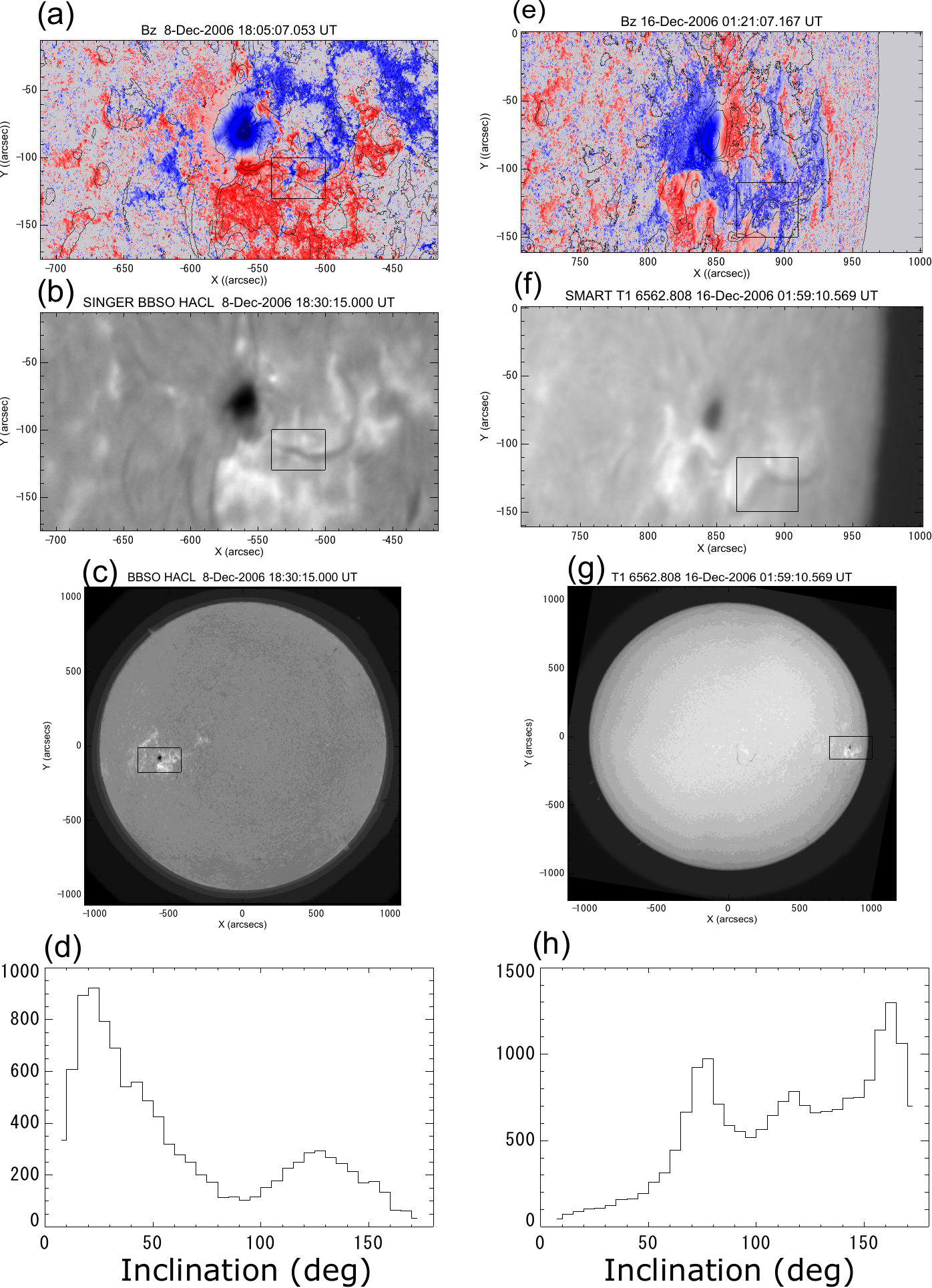}
\end{center}
\caption{
Observations of active region NOAA 10930 (a-d) near the
eastern limb on 2006 December 8 and (e-h) near the western limb on 2006 December 16.
Color maps in panels (a) and (e) show  the line-of-sight component 
of the magnetic field derived by the Milne-Eddington fitting 
from the polarimetric observations by {\it Hinode} SOT/SP. 
The color plot is saturated at an absolute strength at 3 kG
with dark blue (red) for negative (positive) field.
The brightness distributions in the H$\alpha$ band are shown 
as  contours in panels (a) and (e), as a gray-scale map in panels 
(b), (c), (f), and (g).
The data on 2006 December 8 is taken at BBSO and that on 2006 December 16 is
by SMART at the Hida Observatory, respectively.
Panels (d) and (h) show the histogram of the inclination angle 
of the magnetic field.
The analyzed areas are shown as boxes in panels (a) and (e), respectively.
}\label{eastwest}
\end{figure}
Figures \ref{eastwest}a-d show  the observation when this active region is 
near the  east limb ($\Phi <0$).
As noted in the beginning of this section, 
we regard the dark structures that appeared in these panels to be identical 
with those in figure \ref{d061211_xysp_filament_f}
by comparing the continuous images (at least, one image per day) 
of the H$\alpha$ observations.
The LOS component $B_z$ of the photospheric field in the solid insets
is found to be dominated by the positive ($B_z >0$, 
$\gamma_B<90\deg$)
below the filament in this location.
On the other hand, it is negative ($B_z <0$, $\gamma_B>90\deg$)
when the filament is near the west limb ($\Phi >0$, figure \ref{eastwest}e-h).
Thus, there is a variation of the sign in the LOS magnetic component $B_z$
consistent with the above formula (\ref{eq050}),
and one can also obtain the indication of $B_\perp<0$. This means that
the photospheric magnetic field below the 
filament has an inverse-polarity structure (see figure \ref{figure_p1}).
Note that $B_\parallel>0$, $B_\Phi>0$ and $B_\Theta<0$ are also 
suggested from $B_\perp<0$.

In terms of the magnetic field $\bs{B}$ outside the PIL,
we expect that the change of sign in the LOS component $B_z$ occurs at
\begin{equation}
\cot\Phi \approx \frac{B_\Phi}{|B_\Phi|}\tan\gamma_{B0},
\label{eq060}
\end{equation}
where $\gamma_{B0}$ is an inclination angle
(see Appendix \ref{secappendix1}).
Figure \ref{d061211_xysp_azinc}b  shows that the
inclination angle 
of the southern negative polarities in the northern 
side is 
$\gamma_{B0}\approx
160$ deg 
(figure \ref{d061211_xysp_azinc}d). 
Therefore, the sign of the LOS component in the southern side should change 
when the filament is located at
$\Phi \approx -70\ \deg$. 
The observation shows the change
occurred at $\Phi \approx -35\ \deg$. 
For the positive polarities in the northern side, 
since $\gamma_{B0}\approx 25\ \deg$ ,
the sign of the LOS component should change when the filament is located  at
$\Phi \approx 65\ \deg$. 
The measurement 
indicates that the change of sign occurred 
at 
$\Phi \approx 50\ \deg$. 
Taking into account the measurement error of
$\approx$ 10 deg (see appendix \ref{secappendix2}) in the inclination angle, 
the observation supports 
the suggestion of an inverse-polarity structure of the
filament field \textbf{for the northern side.
The quantitative discrepancy for the southern side is difficult to explain.
One possibility is that the magnetic structure of the filament 
gradually changes during the passage which we ignore in the current analysis.
However, there appeared a change of sign in the LOS component even in the
southern side during the passage, which
supports the inverse-polarity structure qualitatively.}

\begin{figure}
\begin{center}
\includegraphics[height=55mm]{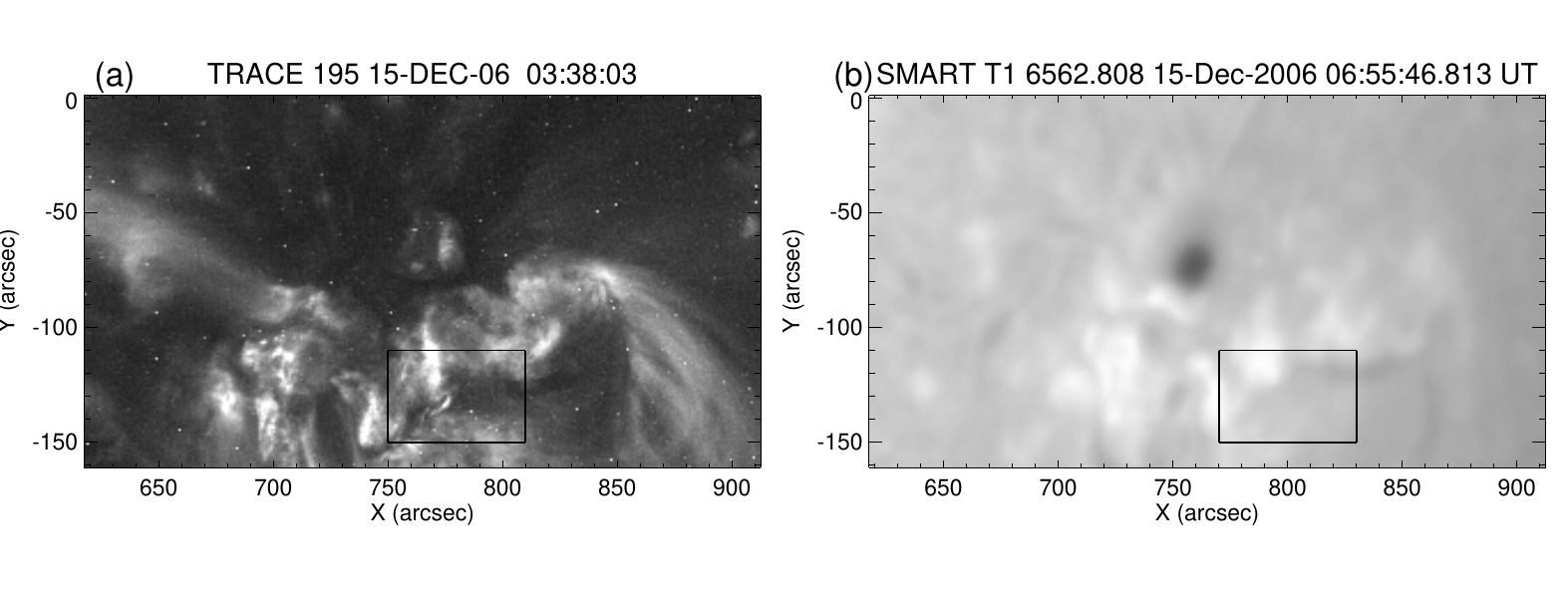}
\includegraphics[height=35mm]{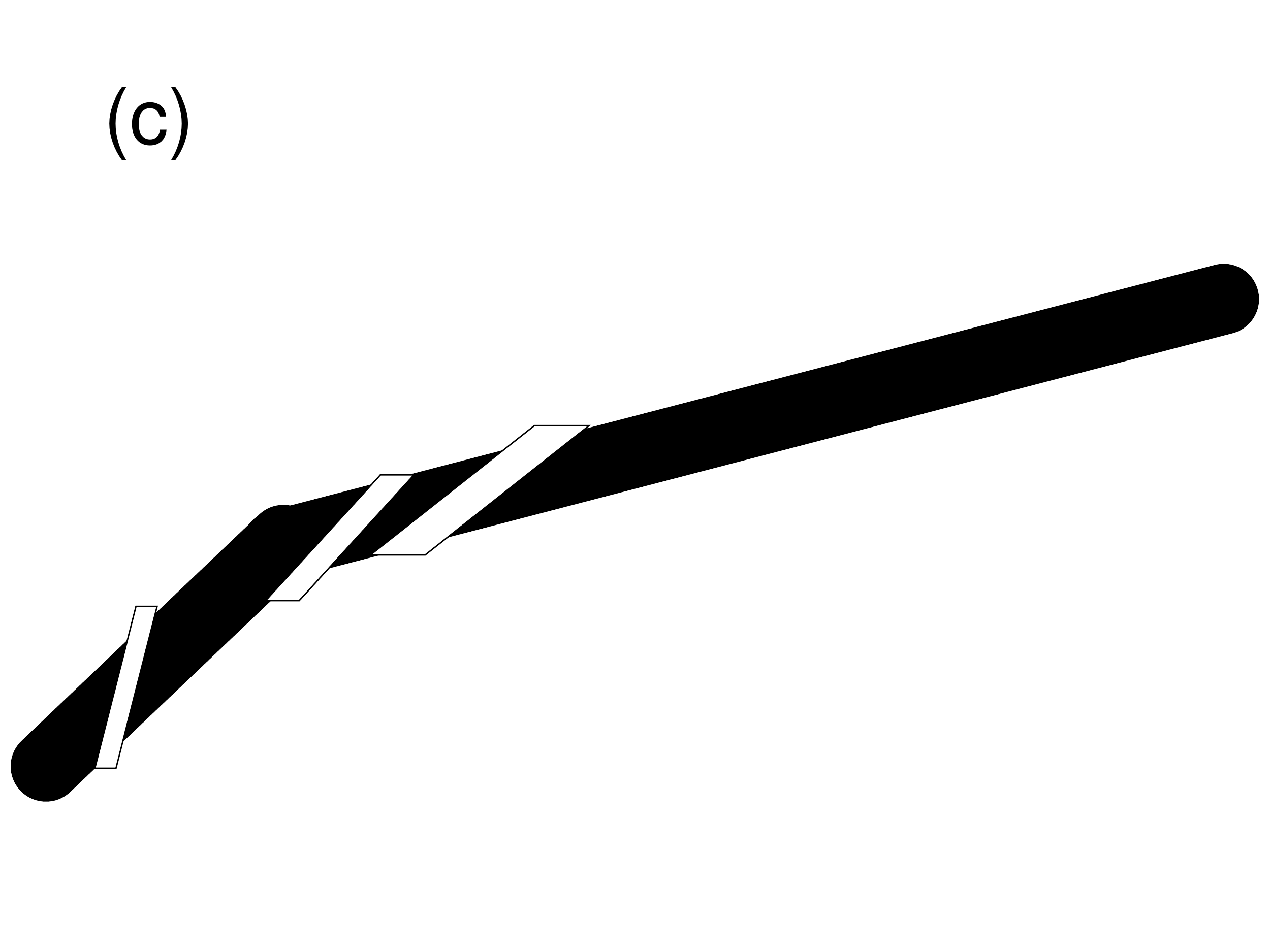}
\includegraphics[height=35mm]{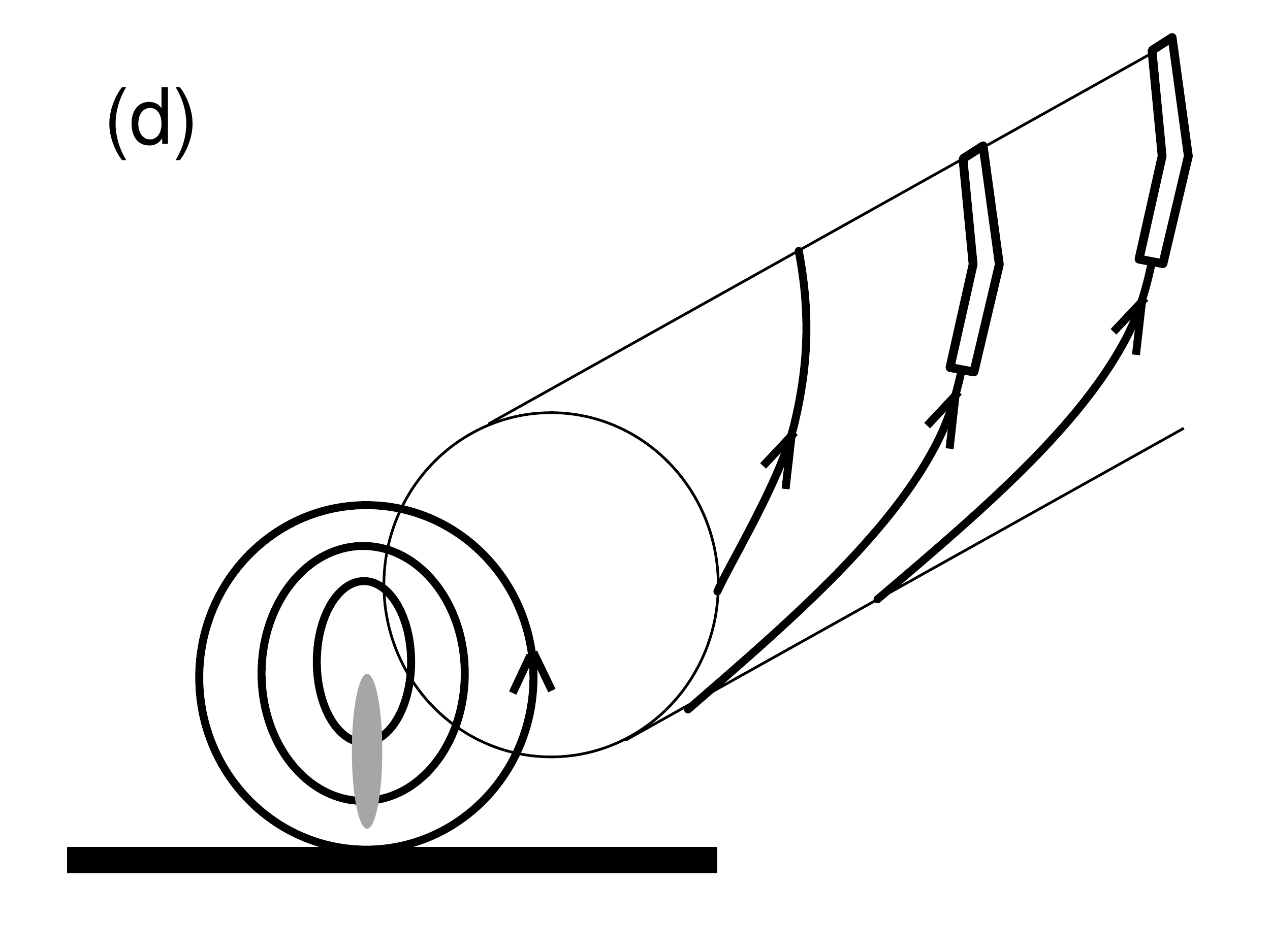}
\end{center}
\caption{
(a) TRACE 195 \AA\ observations of active region NOAA 10930 
at 3:38 UT on 2006 December 15. (b) 
The brightness distributions in the H$\alpha$ band at 6:55 UT on the same date.
The analyzed filament is shown by a solid line box. 
(c) Drawing to indicate the brightening above the dark filament
in the TRACE image in panel (a). (d) A cartoon describing
our explanation of the brightening feature.
}\label{traceeuv}
\end{figure}

The photospheric magnetic field below the filament is indicated 
to have an inverse-polarity. This means that near the PIL,
the magnetic fields have a bald-patch structure, that is, the field lines
are in contact with the photospheric surface in a concave-up shape 
\citep{lites2005}.
Although the global three-dimensional structure high-up in the atmosphere
is out of the scope of this study, 
we have another supporting observational material
for the flux rope interpretation of this filament.
The {\it Transition Region and Coronal Explorer} ({\it TRACE}) 
EUV image in figure \ref{traceeuv} shows
bright features crossing {\it above} the filament in a southeast to 
northwest direction. 
This suggests that the overlying field is twisting the filament in 
a left-handed manner. This is consistent with the orientation of the transverse
component of the photospheric field and the flux-rope interpretation.

The chirality 
\citep{martin1998}
of this filament is also investigated.
Our results show that the axial spine field of the filament is in the westward
direction. According to Martin's definition, this filament has a dextral 
structure (See Fig. 10 of 
\cite{martin1998}).
The detailed structure of the filament material or the surrounding fibrils are
not clear owing to the insufficient spatial resolutions in the available H$\alpha$ images.
The EUV image in figure \ref{traceeuv}a suggests
that the coronal arcades are left-skewed, in a way that is consistent
with the suggestions for the chirality by 
\citet{martin1998}.
\citet{lopezariste2006}
proposed a method to solve the 180-degree ambiguity in
the azimuth angle of the magnetic field by using the chirality of the filament.
Our approach is different from theirs
in that the chirality is not necessary to be known beforehand. It is obtained
as a result of the solution of the ambiguity by using the information on the
LOS component of the photospheric axial field when the filament is located near the limb.

The method of the current analysis includes
several assumptions regarding the filament condition:
There should be no drastic change in the filament magnetic field during the
passage from the east to west limb through the solar disk. If it is dynamic
in nature, it is difficult to interpret the change of sign 
in the LOS component
below the filament as a consequence of the viewpoint or change of structure.
However, this method may add a new approach for studying 
a global magnetic structure
of filaments. As long as stable polarimetric observational data are available,
the orientation of the photospheric axial field can be obtained without 
high resolution 
H$\alpha$ data. This method may be especially promising 
for statistical studies.

\section{Conclusions}

This study focused on the {\it Hinode} observation of the photospheric magnetic 
structure below a filament and attempted to infer what magnetic 
structure the filament has in the equilibrium. We studied a filament in
active region NOAA 10930 that appeared on the solar disk in 2006 December.
The magnetic transverse component perpendicular to the LOS
has a direction almost parallel
to the filament spine with a shear angle of 30 deg, whose orientation
includes the 180-degree ambiguity. We used the center-to-limb variations
for the solution of this ambiguity. When the filament is near the east limb,
we found that the line-of-site magnetic component below is positive,
while it is negative near the west limb. 
This change of sign indicates 
that the horizontal photospheric field perpendicular to the 
polarity inversion line beneath the filament 
has an ``inverse-polarity'', which indicates a flux-rope structure
of the filament supporting field.

{\it Hinode} is a Japanese mission developed and launched by ISAS/JAXA, 
with NAOJ as domestic partner and NASA and STFC (UK) as international partners. 
It is operated by these agencies in co-operation with ESA and NSC (Norway). 
The Global High Resolution H$\alpha$ Network is 
operated by the Big Bear Solar Observatory, New Jersey Institute of Technology 
in U. S. A.  TRACE is a NASA Small Explorer mission.
The authors are supported by JSPS KAKENHI Grant:
T.Y. is by JP15H03640, M.S. by JP17K05397,
Y. K. by JP18H05234 (PI: Y. K.) and JP25220703 (PI: S. Tsuneta),
and JP15H05814 (PI: K. Ichimoto).

\appendix
\section{Derivation of equations (\ref{eq050}) and (\ref{eq060})}
\label{secappendix1}

In this study, we use the heliocentric Cartesian coordinate with
$x$, $y$, and $z$. The
origin is at the solar center, and $x$ and $y$ are, respectively,
in the westward and northward directions 
and $z$ is in the earthward direction.
Further, we use heliographic coordinates $r$, $\Theta$, and $\Phi$, where
$\Theta$ is the latitude and $\Phi$ is the longitude.
The relations among the elemental
unit vectors of these two coordinate systems are given as
\begin{equation}
\begin{array}{cl}
\bs{e}_r &=
 \bs{e}_x \cos{\Theta}\sin{\Phi}
+\bs{e}_y \sin{\Theta}
+\bs{e}_z \cos{\Theta}\cos{\Phi},
\\
\bs{e}_\Phi &=
 \bs{e}_x             \cos{\Phi}
-\bs{e}_z             \sin{\Phi},
\\
\bs{e}_\Theta &=
-\bs{e}_x \sin{\Theta}\sin{\Phi}
+\bs{e}_y \cos{\Theta}
-\bs{e}_z \sin{\Theta}\cos{\Phi}.
\end{array}
\end{equation}

Assume that the spine of the filament is parallel to the PIL
on the surface of the sun and take an elemental vector 
$\bs{e}_\parallel$ to be parallel to the PIL (figure \ref{figure_p1}).  Then,
\begin{equation}
\begin{array}{cl}
\bs{e}_\parallel&=\bs{e}_\Phi \cos{\bar{\chi}_f}+\bs{e}_\Theta \sin{\bar{\chi}_f},
\\
\bs{e}_\perp    &=-\bs{e}_\Phi \sin{\bar{\chi}_f}+\bs{e}_\Theta \cos{\bar{\chi}_f},
\end{array}
\end{equation}
where $\bar{\chi}_f$ is the tilt angle of the PIL against the latitudinal line.
The observed tilt angle ${\chi}_f$,
taking into account of the foreshortening effect, is given as 
\begin{equation}
\tan{{\chi}_f} = \frac{\bs{e}_\parallel\cdot\bs{e}_y}
                    {\bs{e}_\parallel\cdot\bs{e}_x}.
\end{equation}
Solving this for $\bar{\chi}_f$, we obtain
\begin{equation}
\tan{\bar{\chi}_f} = \frac{\cos{\Phi}}
{\sin{\Theta}\sin{\Phi}+\cos{\Theta}/\tan{{\chi}_f}}.
\label{eqa40}
\end{equation}
The magnetic field component
perpendicular to the PIL in the solar surface 
is given as
\begin{equation}
B_\perp = \bs{B}\cdot \bs{e}_\perp
        =  B_\Theta \cos{\bar{\chi}_f}
        - B_\Phi   \sin{\bar{\chi}_f}.
\label{eqa70}
\end{equation}
The azimuth $\bar{\chi}_B$ is defined as
\begin{equation}
\tan \bar{\chi}_B=\frac{B_\Theta}{B_\Phi}.
\end{equation}
The relations between components are
\begin{equation}
\begin{array}{cl}
B_z&=
 B_r      \cos{\Theta}\cos{\Phi}
-B_\Phi               \sin{\Phi}
-B_\Theta \sin{\Theta}\cos{\Phi},
\\
B_x  &=
 B_r      \cos{\Theta}\sin{\Phi}
+B_\Phi               \cos{\Phi}
-B_\Theta \sin{\Theta}\sin{\Phi},
\\
B_y &=
 B_r      \sin{\Theta}
+B_\Theta \cos{\Theta},
\end{array}
\label{eqa50}
\end{equation}
and their observed azimuth angle is
\begin{equation}
\tan\chi_B =\frac{B_y}{B_x}.
\label{eqa60}
\end{equation}

In terms of the magnetic field $\bs{B}$ on the PIL, 
the local vertical component to the surface is zero, i.e.,
\begin{equation}
B_r=0.
\label{eqa80}
\end{equation}
By using eqs. (\ref{eqa50}) and (\ref{eqa60}) with (\ref{eqa80}), we obtain
\begin{equation}
\tan{\bar{\chi}_B} = \frac{\cos{\Phi}}
{\sin{\Theta}\sin{\Phi}+\cos{\Theta}/\tan{{\chi}_B}}.
\label{eqa90}
\end{equation}
We are interested in the relation between the LOS component $B_z$
and the perpendicular component $B_\perp$
to the PIL.
Using eqs. (\ref{eqa50}) and (\ref{eqa70}) with (\ref{eqa80}),
their ratio is given as
\begin{equation}
\frac{B_z}{B_\perp}
=\frac{ -\sin{\Phi}-\sin{\Theta}\cos{\Phi}\tan{\bar{\chi}_B}}
{ \tan\bar{\chi}_B\cos{\bar{\chi}_f} -\sin{\bar{\chi}_f} }.
\label{eqa110}
\end{equation}

Suppose the analyzing filament is close to the equator, i.e., 
$|\Theta| \ll 1 \ \rad$ like 
our analyzed one. 
Then, from eq. (\ref{eqa40}),
\begin{equation}
\tan{\bar{\chi}_f} \approx
\cos{\Phi} \tan{{\chi}_f},
\end{equation}
and from eq. (\ref{eqa90}),
\begin{equation}
\tan{\bar{\chi}_B} \approx
\cos{\Phi}  \tan {{\chi}_B}.
\end{equation}
To obtain 
$\bar{\chi}_B$ and $\bar{\chi}_f$, the procedure 
becomes simpler by using data at $|\Phi| \ll 1 \ \rad$ as
\begin{equation}
{\bar{\chi}}_f \approx {\chi}_{f0}
\label{eqa140}
\end{equation}
and
\begin{equation}
{\bar{\chi}}_B \approx {\chi}_{B0}.
\label{eqa150}
\end{equation}
Here, the subscript $0$ denotes a value at $\Phi \approx 0 \ \rad$.
We hereafter assume that these values do not change during the disk passage.
Since the observed tilt angle of the filament is small, i.e.,
$|\bar{\chi}_{f}| \approx |\bar{\chi}_{f0}| \ll 1 \ \rad$,
from eq. (\ref{eqa110}), we obtain
\begin{equation}
\frac{B_z}{B_\perp}
\approx 
-\frac{\sin{\Phi}}{\tan{\bar{\chi}_B}}
-\sin\Theta \cos\Phi.
\end{equation}
Since the observed azimuth angle of the transverse field is small, i.e.,
$|\bar{\chi}_{B}| \approx |\bar{\chi}_{B0}| \ll 1 \ \rad$,
we finally obtain eq. 
(\ref{eq050}) as
\begin{equation}
\frac{B_z}{B_\perp}
\approx 
-\frac{\sin{\Phi}}{\tan{{\chi}_{B0}}}.
\end{equation}

The observed inclination angle is given as
\begin{equation}
\tan\gamma_B =\frac{\sqrt{B_x^2+B_y^2}}{B_z}.
\label{eqa65}
\end{equation}
In terms of the magnetic field $\bs{B}$ outside the PIL,
using $|\Theta| \ll 1 \ \rad$,
\begin{equation}
\begin{array}{cl}
B_z&\approx B_r      \cos{\Phi}
        -B_\Phi      \sin{\Phi},
\\
B_x  &\approx B_r    \sin{\Phi}
        +B_\Phi      \cos{\Phi},
\\
B_y &\approx 0,
\end{array}
\end{equation}
and
\begin{equation}
\tan\gamma_B \approx \frac
{|B_r \sin{\Phi} + B_\Phi \cos{\Phi}|}
{ B_r \cos{\Phi} - B_\Phi \sin{\Phi} }.
\end{equation}
If one uses the data at $|\Phi| \ll  0 \ \rad$ then
\begin{equation}
\tan\gamma_{B0} \approx 
\frac {|B_\Phi|}{ B_r } .
\end{equation}
Note that this value is assumed to remain constant during the disk passage.
The LOS component is
\begin{equation}
B_z \approx |B_\Phi| 
\left(
\frac{\cos\Phi}{\tan\gamma_{B0}}-\frac{B_\Phi}{|B_\Phi|}\sin\Phi
\right),
\end{equation}
and its change of sign occurs at
\begin{equation}
\cot\Phi \approx \frac{B_\Phi}{|B_\Phi|}\tan\gamma_{B0},
\end{equation}
which is (\ref{eq060}).

\section{MEKSY code}
\label{secappendix2}

The fitting code MEKSY is developed by the 
authors of this paper.
It is based on MELANIE 
\citep{socas2008}
for the fitting with the model atmosphere
and PIKAIA 
\citep{charbonneau2002}
for the initial guess.
The observed Stokes profiles are fitted by a spectral model
to obtain the physical parameters in the
solar photosphere. The model is based on the Milne-Eddington
atmosphere in which the physical parameters, such as magnetic field
strength, orientation, and Doppler velocity, are assumed to be homogeneous
along the line of sight. An exact analytical solution is available
as a set of explicit formulae known as the Unno-Rachkovsky solution.
The details of this solution are described, e.g., by
\citet{deltoroiniesta2003}.

The fitting parameters are:
(1) magnetic field strength, 
(2) magnetic field inclination,
(3) magnetic field azimuth,
(4) line strength (the opacity ratio of the line to the continuum),
(5) Doppler width,
(6) ratio of the damping width and Doppler width,
(7) Doppler velocity,
(8) radiative source function,
(9) gradient of the radiative source function,
(10) macro-turbulence velocity,
(11) stray-light fraction, and
(12) wavelength shift of the stray-light component.
The model profile is a blend of the polarized and
the non-polarized components. The former is the radiation from the
magnetized plasma in the solar atmosphere, and the latter is
a mixture of the radiation from the non-magnetized plasma and
that from the stray-light in the instrument.
The polarized component is synthesized based on
the Unno-Rachkovsky formula using the above parameters, whereas
the stray-light component is extracted as the
average of the observed profiles over the pixels with a weak polarization degree
in the corresponding scan map.
The standard Levenberg-Marquardt (LM) method (e.g. 
\cite{press1992})
is used for the non-linear fitting algorithm. This part of MEKSY is
a tune-up version of MELANIE. Through the development of the
code, we found that the quality of the convergence
of the iterative procedures in the LM fitting is highly dependent on
the initial-guess parameter sets. In MEKSY, they are provided by
the PIKAIA code which is based on the genetic algorithm
\citep{charbonneau1995}.
For data input and output,
the CFITSIO library 
\citep{william1999}
is used.

The main purpose behind the development of the MEKSY code is the fulfillment of
high-speed processing. One {\it Hinode} SP normal-mode map includes 
approximately a square of one kilo pixel points. This is an
enormous number compared with the previous spectro polarimetric
data, e.g. the Advance Stokes Polarimeter typically has only 256$^2$
pixel points. {\it Hinode} also observes the Sun 
with practically no interruption and may routinely
produce large format data. The developed code is optimized as much as possible 
to meet these requirements. An IDL front-end 
macro program is also developed to submit parallel processes
on the parallel computing system (grid engine system).
As a result, the process time is 50 mili second for each pixel
by an Opteron 2.2-GHz core, and thus it takes
14 hours to build a $\approx 1000^2$-pixel map. The process can be parallelized
by using the parallel system so that the entire process takes 
less than 1 hour by using a 16-CPU system. 
The code is installed on the grid engine system
at the Hinode Science Center (HSC) in the National Astronomical Observatory of 
Japan\footnote{The system started its operation from 2006 and was shutdown in 
February 2013. A similar system is in operation as a part of the 
multi-wavelength data analysis system of the Astronomy Data Center, NAOJ.}.
The program source is written in Fortran 90 and
the technical details of the code
including the user reference manual are available on the HSC website 
as online documents
\footnote{http://hinode.nao.ac.jp/SDAS/SSW-IDL\_on\_ADC\_E.shtml}.

\begin{figure}
\begin{center}
\includegraphics[width=130mm]{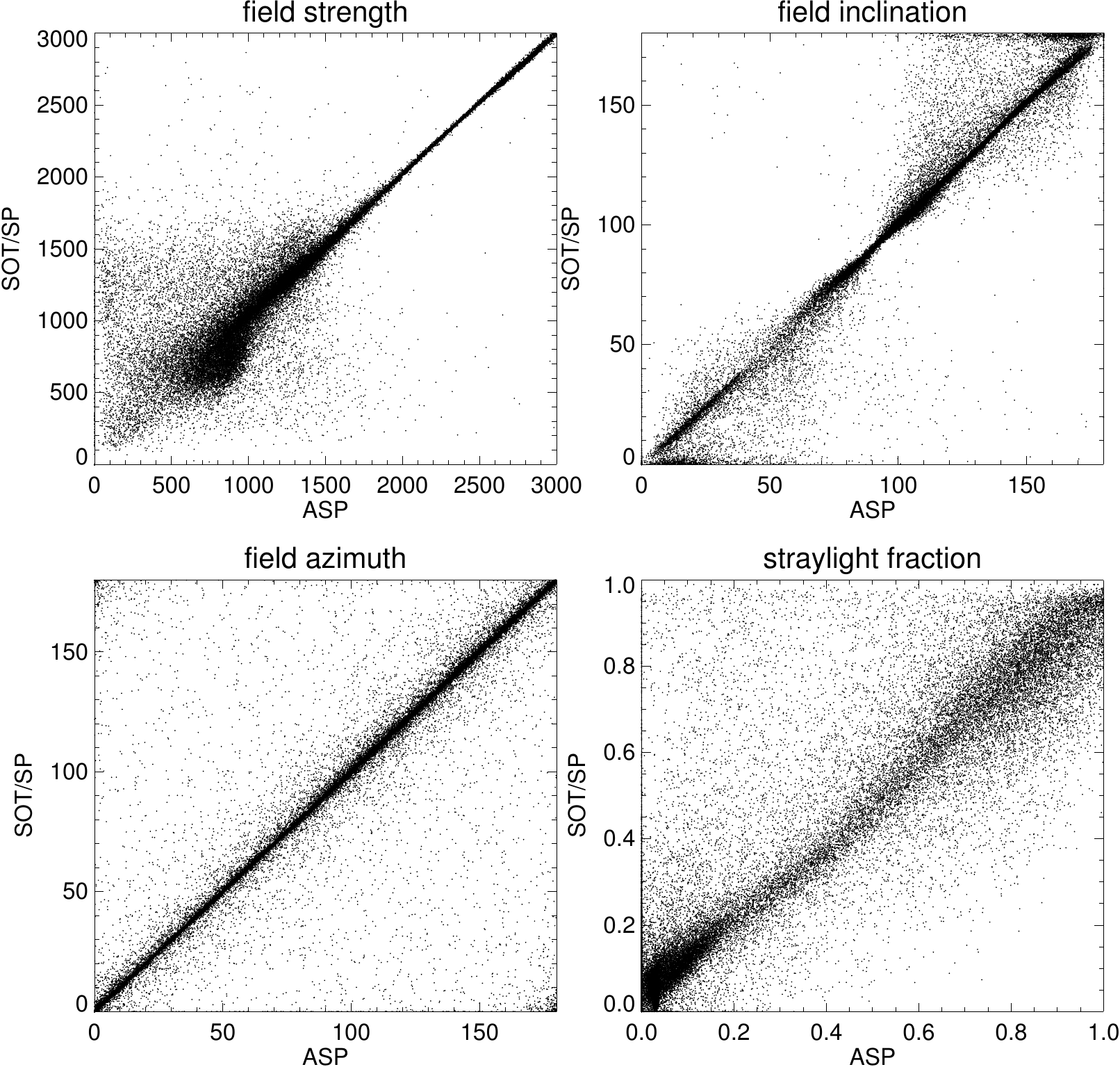}
\includegraphics[width=65mm]{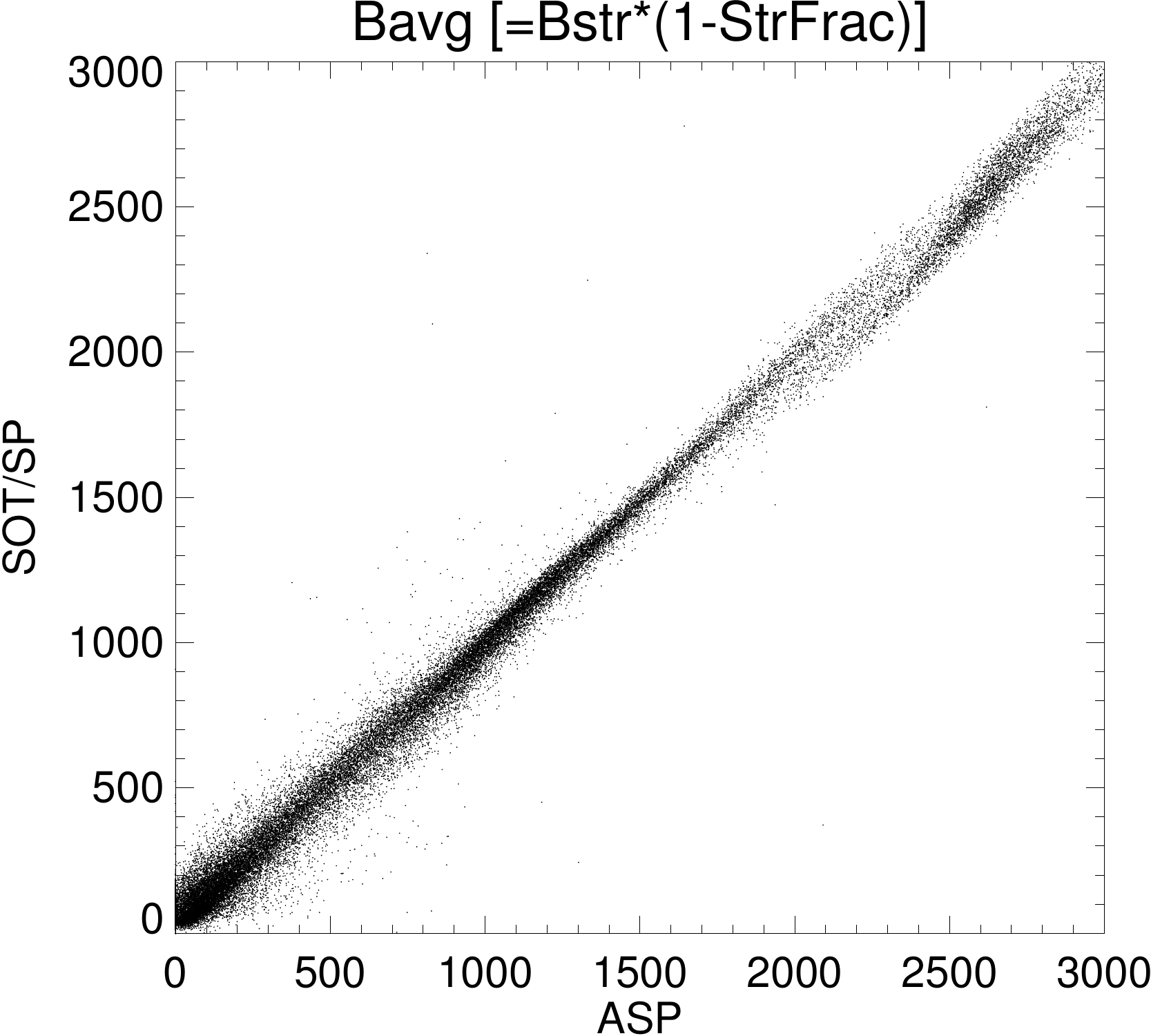}
\end{center}
\caption{
Scatter plots of the results on the different inversion codes. The vertical
axis represents the results obtained by MEKSY, and the horizontal 
axis represents those obtained by the ASP code.
Each point corresponds to each spatial pixel. (upper left) Field strength
in Gauss, (upper right) field inclination in degrees, 
(middle left) field azimuth in degrees, (middle right) stray-light fraction
(a complement of the filling factor), and (bottom) magnetic flux density
(i.e., the product of the filling factor and field strength).
}\label{meksyaspcomp}
\end{figure}

\begin{figure}
\begin{center}
\includegraphics[width=160mm]{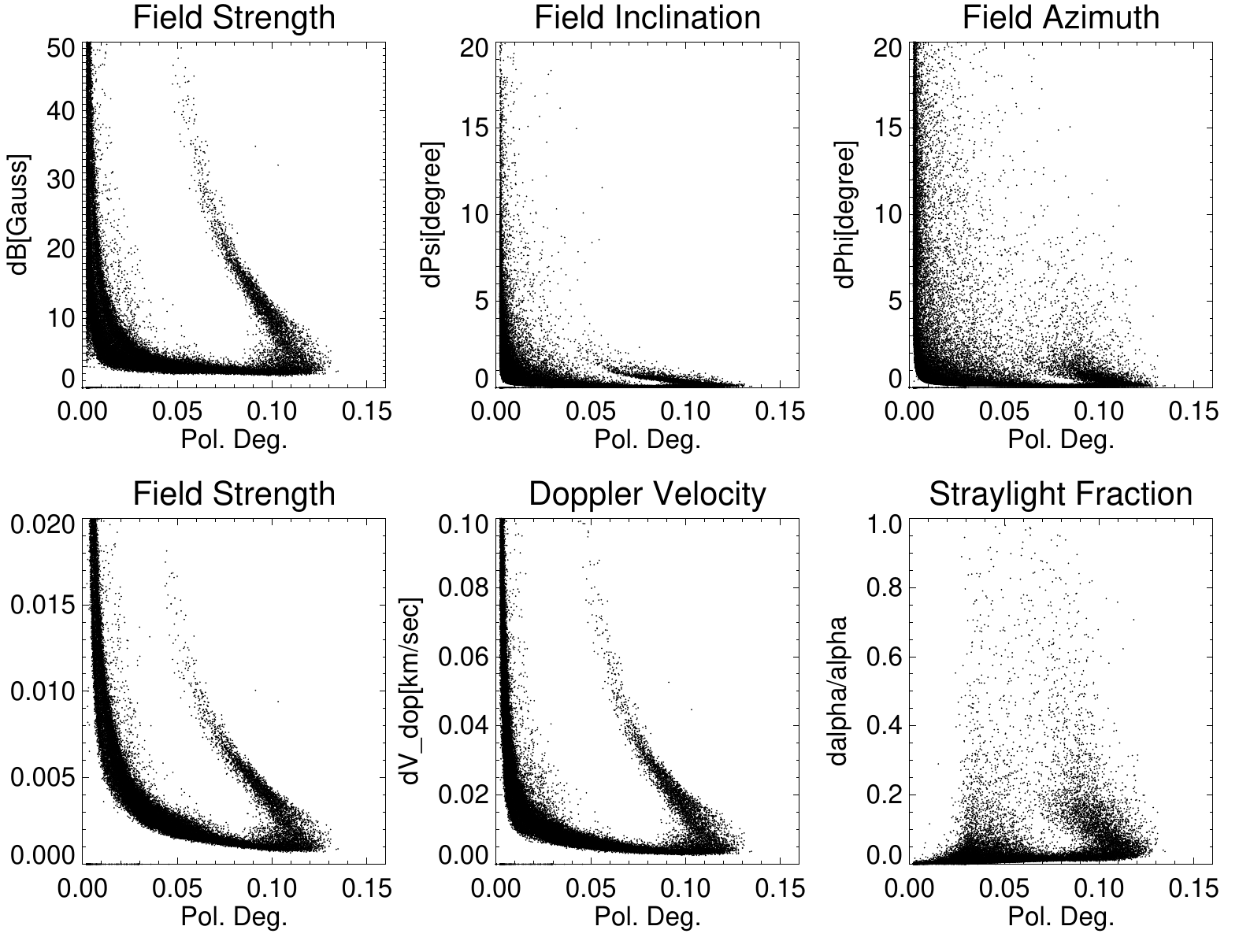}
\end{center}
\caption{
Derivation errors as functions of the polarization degrees in each pixel.
(Top left) field strength in Gauss, (top center)  field inclination in degrees,
(top right) field azimuth in degrees, (bottom left) the ratio of 
the error in field 
strength to the amount, (bottom center) Doppler velocity, (bottom right)
the ratio of the error in stray-light fraction to the amount.
}\label{meksyerror}
\end{figure}

Figure \ref{meksyaspcomp} shows the comparison of the fitting results with the
existing code. We used the results of the so-called ``ASP code''
\citep{skumanich1987, lites1994}
as a reference developed 
for the inversion of the Stokes data obtained by the
Advanced Stokes Polarimeter (ASP). The results are almost consistent.
Although there is crosstalk between the field strength and the stray-light
fraction when the strength is weak, the average magnetic flux density
is consistent (bottom panel of figure \ref{meksyaspcomp}). 

Figure \ref{meksyerror} shows the errors of the fitting
as a function of the polarization degree in each spatial pixel.
Note the data contains the image of the sunspot umbra where the Stokes
profile has complex figure presumably because of 
the effect of the low-temperature
molecular lines that are not taken into account in the present code.
The branches extending in the relatively strong 
($\approx 0.1$) polarization degree
range are attributed to this effect. Otherwise, we may estimate the error of the
fitting procedure by using MEKSY. The one-sigma errors in the pixels
with polarization degrees greater than 2 \% are: a few Gauss for 
field strength, a few degrees both for the field inclination and 
field azimuth, $\approx 10\ {\rm m/sec}$ for the Doppler velocity, and 
$\approx$ 10 \% for the stray-light fraction.
The error for each variable is 
calculated as $\delta a_{k}=(\chi^{2} / \alpha_{kk})^{1/2}$, where
$a_{k}$ ($k=$1, ..., 12) 
is the fitting parameter given above, and $\delta a_{k}$
is its fitting error, $\chi^{2}$ is the value of the merit function for the obtained
fit results, and $\alpha_{kk}$ is the $k$-th diagonal component of the ``curvature matrix''
defined as the curvature of the merit function $\chi^2(a_k)$ in the fitting parameter space.
The detailed definitions of these variables are given in 
\citet{press1992}.

\bibliography{draftref}

{}

\end{document}